\pdfoutput=1
\documentclass[letterpaper,english,aps,letter,superscriptaddress,nofootinbib,floatfix,pre,twocolumn]{revtex4}
\usepackage[T1]{fontenc}
\usepackage[latin1]{inputenc}
\usepackage{graphicx}
\usepackage{amssymb}

\makeatletter

\providecommand{\boldsymbol}[1]{\mbox{\boldmath $#1$}}

\usepackage{float}
\floatstyle{ruled}
\newfloat{algorithm}{tbp}{loa}
\floatname{algorithm}{Algorithm}

\usepackage{ae,aecompl}

\renewcommand{\citet}[1]{\cite{#1}}

\newenvironment{Algorithm}[1] {
  \refstepcounter{algorithm} 
  \vspace{1ex} \hrule \vspace{1ex} 
  \center{\textbf{Algorithm \thealgorithm : }#1}
  \vspace{1ex} \hrule \vspace{1ex}
  \small
} 
{ \normalsize \vspace{1ex} \hrule \vspace{1ex} }

\usepackage{babel}
\makeatother
\begin{document}

\title{Asynchronous Event-Driven Particle Algorithms}

\author{Aleksandar Donev}

\affiliation{Address: Lawrence Livermore National Laboratory, P.O.Box 808, L-367,
Livermore, CA 94551-9900\\
Phone: (925) 424-6816, Fax: (925) 423-0785, E-mail: \emph{aleks.donev@gmail.com}}

\begin{abstract}
We present, in a unifying way, the main components of three asynchronous
event-driven algorithms for simulating physical systems of interacting
particles. The first example, hard-particle molecular dynamics, is
well-known. We also present a recently-developed diffusion kinetic
Monte Carlo algorithm, as well as a novel stochastic molecular-dynamics
algorithm that builds on the Direct Simulation Monte Carlo. We explain
how to effectively combine asynchronous event-driven with classical
time-driven or with synchronous event-driven handling. Finally, we
discuss some promises and challenges for event-driven simulation of
realistic physical systems.

\end{abstract}
\maketitle
\newcommand{\Cross}[1]{\left|\mathbf{#1}\right|_{\times}}
\newcommand{\CrossL}[1]{\left|\mathbf{#1}\right|_{\times}^{L}}
\newcommand{\CrossR}[1]{\left|\mathbf{#1}\right|_{\times}^{R}}
\newcommand{\CrossS}[1]{\left|\mathbf{#1}\right|_{\boxtimes}}

\newcommand{\V}[1]{\mathbf{#1}}
\newcommand{\M}[1]{\mathbf{#1}}
\newcommand{\D}[1]{\Delta#1}

\newcommand{\sV}[1]{\boldsymbol{#1}}
\newcommand{\sM}[1]{\boldsymbol{#1}}

\newcommand{\grad}{\boldsymbol{\nabla}}
\newcommand{\eij}{\left\{  i,j\right\}  }

\section{Introduction}

There is a wide range of particle systems from computational science
problems that can be efficiently and accurately simulated using \emph{asynchronous
event-driven} (AED) algorithms. Two prominent examples are \emph{molecular
dynamics} (MD) \citet{EventDriven_Alder} for systems of hard particles
such as disordered granular packings \citet{Event_Driven_HE}, polymers
\citet{EDMD_Polymers_Hall}, colloids \citet{EventDriven_Colloids},
particle-laden flows \citet{EventDriven_ParticleField}, and others,
as well as \emph{kinetic Monte Carlo} (KMC) \citet{AdvancedKMC_Tutorial}
for diffusion-limited reactions \citet{FPKMC_PRL}, epitaxial growth
\citet{KMC_SurfaceReactions}, quantum systems \citet{QKMC_GreensFunction},
bio-chemical reaction networks \citet{ParticleBased_CellModeling},
self-assembly \citet{EventDriven_BioSelfAssembly}, and many others.
As of yet unexplored are \emph{multi-scale} and \emph{multi-physics}
algorithms such as event-driven dislocation dynamics, phase separation/precipitation,
combined flow and diffusion with (bio)chemical reactions, etc.

In this work we will focus on a class of particle-based problems that
are very common in computational materials science and are well-suited
for AED simulation. Specifically, we will focus on the simulation
of large systems of mobile particles interacting with \emph{short-range
pairwise} (two-body) potentials (forces). Our goal will be to reveal
the common building blocks of these simulations (e.g., event queues,
neighbor searches), but also to highlight the components that are
problem specific (e.g., event prediction and processing). We will
present these components in some detail for three specific examples:
event-driven molecular dynamics (EDMD), first-passage kinetic Monte
Carlo (FPKMC), and stochastic EDMD (SEDMD). Through the discussion
of these examples we will demonstrate the undeniable advantages of
AED algorithms, but we will also reveal the difficulties with using
AED algorithms for realistic models.

\subsection{Background}

We consider the simulation of the time evolution of a collection of
$N$ interacting particles in $d$-dimensions, starting from some
initial condition. At any point in time, the system $\mathcal{Q}=\left(\V{Q},\mathcal{B}\right)$
is characterized by the \emph{configuration} $\V{Q}=\left(\V{q}_{1},\ldots,\V{q}_{N}\right)$,
containing at least the centroid positions $\V{r}_{i}$ for every
particle $i$, and the additional global information $\mathcal{B}$,
which may involve variables such as boundary conditions or external
fields. The number of particles $N$ may itself vary with time. For
each particle $i$ we may consider an arbitrary number of attributes
$\V{a}_{i}$ in addition to the position of the centroid $\V{r}_{i}$,
$\V{q}_{i}=\left(\V{r}_{i},\V{a}_{i}\right)$, for example, $\V{a}_{i}$
may also contain the linear and/or angular velocity, the orientation
and/or the chemical species (shape, charge, mass, internal composition)
of particle $i$. Typically $\V{a}_{i}$ will at least contain an
integer that identifies its \emph{species} $1\leq s_{i}\leq N_{s}$,
and some information will be shared among all particles belonging
to the same species (ex., the charge or mass). In particular, the
symmetric \emph{interaction table} $\mathcal{I}_{\alpha\beta}$ stores
$N_{s}(N_{s}+1)/2$ logical (true or false) entries that specify whether
species $\alpha$ and $\beta$ interact or not.

Two particles $i$ and $j$ are \emph{overlapping} only if a certain
(generalized) distance between them $d_{ij}(\V{q}_{i},\V{q}_{j})\geq0$
is less than some \emph{cutoff distance} or \emph{diameter} $D_{ij}$.
Overlapping particles react with each other in an application specific
manner. Typically the type of reaction and $d_{ij}$ depend only on
the species of the two particles, but there may also be dependencies
on time or some other external field parameters. For example, for
(additive) hard spheres $d_{ij}=\left\Vert \V{r}_{i}-\V{r}_{j}\right\Vert _{L_{2}}$
is the $L_{2}$ norm of $\V{r}_{ij}=\V{r}_{i}-\V{r}_{j}$, and $D_{ij}=(D_{i}+D_{j})/2$
and the type of reaction is (hard-core) repulsion. For a non-interacting
pair of particles one may set $D_{ij}<0$. Particles may also overlap
with boundaries of the simulation domain, such as hard walls or reactive
surfaces, however, typically the majority of interactions are among
particles. Our assumption of short-range interactions implies that
all $D_{ij}$'s are much smaller than the system size. If there are
long range interactions present (e.g., electrostatic forces in plasmas),
it may sometimes be possible to split them into a short-range part
and a long-range part and treat the long-range part separately (e.g.,
multipole or cell-based FFT methods) as part of the {}``boundary''
handling.

We will assume that the time evolution (motion) of the system is mostly
smooth with the exception of certain discrete \emph{events}, which
lead to discontinuous changes in the configuration of a certain collection
of particles. Events may involve a single particle $i$, such as the
change of the internal state of the particle (e.g., decay reactions,
spin flips, sudden changes in the particle velocity). Events may also
involve pairs of particles, for example, the \emph{collision} (exchange
of momentum) between two particles $i$ and $j$, or a chemical reaction
between two overlapping particles $i$ and $j$ leading to the creation
and/or destruction of particles. For now we will ignore events involving
more than two particles. Some events may also involve global variables
in $\mathcal{B}$ and thus implicitly affect all of the particles.
Here we will assume events are instantaneous, i.e., they have no duration.
Events that have a duration (e.g., particles overlapping for a certain
time interval) can be treated as a pair of events, one for the start
of the event and one for the end of the event. In a more general framework
one may consider all events as having a finite duration, as in process-oriented
simulations \citet{ProcessOrientedSimulation} (in such a framework
collisions would be a special degenerate case of a more general {}``overlap''
event).

We will refer to simulations as \emph{event-driven} (also called \emph{discrete
event simulations} \citet{TimeWarp_Book}) if the state of the system
is evolved discontinuously in time from one event to another, \emph{predicting}
the time of the next event whenever an event is processed. This is
in contrast with the more common \emph{time-driven} simulations, where
the state changes continuously in time and is evolved over a sequence
of small time steps $\D{t}$, discovering and processing events at
the end of the time step. Time-driven algorithms are inaccurate when
the time step is large, and they become asymptotically accurate as
$\D{t}\rightarrow0$. Therefore, the time step must be smaller than
the estimated fastest time scales in the problem. This leads to large
inefficiencies when there are multiple dynamically-changing time scales.
On the other hand, event-driven algorithms automatically adjust the
time step.

We will focus on \emph{asynchronous} event-driven (AED) algorithms.
In asynchronous algorithms, there is a global simulation time $t$,
typically the time when the last processed event occurred, and each
particle is at a different point in time $t_{i}\leq t$, typically
the last time it participated in an event. This is to be contrasted
to synchronous event-driven algorithms, where all of the particles
are at the same time $t$. One of the most important examples of a
synchronous event-driven algorithm in materials science is the $n$-fold
(BKL) algorithm for performing kinetic (dynamic) Monte Carlo simulations
\citet{AdvancedKMC_Tutorial}. The exactness and efficiency of the
$n$-fold algorithm hinge on the fact that the state of the system
does not change in-between events, as is common in lattice models
where the positions of the particles are discrete. For example, atoms
may stay in the immediate vicinity of the crystal lattice sites and
only sometimes hop to nearby sites. In the types of problems that
we will consider, the positions of the particles will be continuous
and continuously changing even in-between events. Therefore, synchronous
KMC algorithms will be approximate and can be viewed as the equivalent
of time stepping, where the time step $\D{t}$ is not constant. It
is important to point out that even in cases where the evolution of
the system consists entirely of discrete jumps (e.g., Markov chain
transitions), asynchronous algorithms may be more efficient than synchronous
ones \citet{EventDriven_BioSelfAssembly}. It is possible to combine
the two algorithms by using the more general asynchronous algorithm
at the top level but treating a subset of the particles synchronously,
as if they are a super-particle with complex internal structure.

\subsection{Model Examples}

Atomistic or molecular-level modeling is one of the foundations of
computational materials science. The two most popular types of algorithms
used in the simulation of materials are molecular dynamics (MD) and
Monte Carlo (MC) algorithms. Monte Carlo algorithms are often used
to study static equilibrium properties of systems, however, here we
focus on dynamic or kinetic Monte Carlo, where the time evolution
of a system is modeled just like in MD. For our purposes, the only
important difference between the two is that MD is a deterministic
algorithm, in which deterministic equations of motions are solved,
and Monte Carlo is a stochastic procedure in which sample paths from
an ensemble of weighted paths are generated. In both cases one typically
averages over multiple trajectories, starting with different initial
conditions and/or using a different random number seed. From the perspective
of AED algorithms, this means that random number generators (RNGs)
are involved in the determination of the time certain events occur
as well as in the actual processing of those events.

The very first \emph{molecular dynamics} (MD) calculations simulated
a system of hard disks and hard spheres and used an AED algorithm
\citet{EventDriven_Alder}. Event-driven MD (EDMD) algorithms for
hard particles are discussed in considerable detail in Ref. \citet{Event_Driven_HE}
and elsewhere, here we only present the essential components. The
hard-sphere system is a collection of non-overlapping spheres (or
disks), contained within a bounded region, each moving with a certain
velocity $\V{v}_{i}=\dot{\V{r}_{i}}$. Pairs of spheres collide and
the colliding particles bounce off elastically, preserving both linear
momentum and energy. Many generalizations can be considered, for example,
the spheres may be growing in size and/or the particles may be nonspherical
\citet{Jammed_MM}, the collisions may not be perfectly elastic \citet{DSMC_Granular},
some of the particles may be tethered to each other to form structures
such as polymer chains \citet{EDMD_Polymers_Hall}, etc. The general
features are that particles move ballistically along simple deterministic
paths (such as straight lines) in-between binary collisions. The primary
type of event are binary collisions, which have no duration and involve
deterministic changes of the velocities of the colliding particles.
The ballistic motion of the particles is described by Newton's equations
of motion (i.e., deterministic ODEs), $m\dot{\V{v}}_{i}=\V{F}_{i}$,
where $\V{F}$ is an external forcing (e.g., gravity). 

\emph{Direct simulation Monte Carlo} (DSMC) \citet{DSMCReview_Garcia}
is an MC algorithm that tries to mimic MD for fluids. We will consider
DSMC as a fast alternative to MD, even though it can also be viewed
as a particle-based MC method for solving the Boltzmann equation in
dilute fluids. From our perspective, DSMC is an approximate variant
of MD in which the particle collisions are not processed exactly,
rather, particle collisions are stochastic and (attempt to) follow
the same probability distributions as would have exact MD. Specifically,
nearby particles are randomly chosen to undergo stochastic collisions
and exchange momentum and energy, thus leading to local conservation
laws and hydrodynamic behavior. DSMC is applicable in cases when the
structure of the fluid and the detailed motions of all of the particles
do not matter, as is the case with solvent molecules (e.g., water)
in fluid flow problems or large-scale granular flows \citet{DSMC_Granular}.
Traditionally DSMC has been implemented using a time-driven approach,
in which at each time step particles are first propagated in a ballistic
(convective) fashion, and then a certain number of stochastic particle
collisions among nearby particles are processed. Here we describe
a novel AED algorithm for DSMC, and demonstrate how it can be integrated
with EDMD in order to replace the expensive MD with cheaper DSMC for
some of the particle species (e.g., solvent molecules). We term the
resulting algorithm Stochastic EDMD (SEDMD).

The motion of the particles in-between events is not always deterministic.
In particular, an important class of problems concerns diffusing particles,
that is, particles whose velocity changes randomly very frequently
(i..e, they make many small steps in random directions). The motion
of the particles is probabilistic, in the sense that the probability
$c(\V{r},t+\D{t})$ of finding a particle at a given position $\V{r}$
at a certain time $t+\D{t}$, assuming it started at the (space and
time) origin, is the solution to the time-dependent diffusion equation
$\partial_{t}c=D\grad^{2}c$ (a deterministic partial differential
equation), where $D$ is the particle diffusion coefficient (generally
a tensor). A variety of \emph{reactions} may occur when a pair of
particles collides, for example, particles may repel each other (colloids
\citet{EventDriven_Colloids}), they may stick or begin merging together
(paint suspensions), or they may undergo a chemical reaction that
consumes the reacting particles and produces zero, one, two, or possibly
more new particles (a wide range of diffusion-limited reactions in
materials \citet{GFRD_KMC,FPKMC_PRL} and biological systems \citet{GFRD_KMC}).
Several \emph{approximate} event-driven KMC algorithms have been used
in the past for this problem \citet{GFRD_KMC}. Here we will describe
a recently-developed First-Passage Kinetic Monte Carlo (FPKMC) \emph{exact}
AED algorithm for simulating a collection of diffusing hard particles
\citet{FPKMC_PRL}. It is worth pointing out that one may also consider
particles whose trajectories are a combination of ballistic and diffusive
motion, that is, motion that is described by Langevin's equations
(or other stochastic ODEs or even PDEs). In that sense, we will see
that both the MD and MC algorithms share many common features.

\section{A General AED Particle Algorithm}

In this work we focus on systems where particles only interact with
nearby particles. We will formalize this by defining a geometric hierarchy
of regions around a given particle. These particle proximity hierarchies
are at the core of \emph{geometry-specific} (GS) aspects of AED simulation,
which can be reused for different \emph{application-specific} (AS)
rules for moving and interacting the particles. We will assign a \emph{hard
core} $\mathcal{C}_{i}$ to each particle such that a particle may
overlap with another particle only if their cores overlap. For (additive)
hard spheres, the core is nothing more than the particle itself. Next,
we \emph{protect} particle $i$ against other particles by enclosing
it inside a \emph{protective region} $\mathcal{P}_{i}$, $\mathcal{C}_{i}\subseteq\mathcal{P}_{i}$,
that is typically disjoint from the majority of other protective regions.
Finally, we assume that every protective region $i$ is contained
within a \emph{neighborhood region} $\mathcal{N}_{i}$, $\mathcal{P}_{i}\subseteq\mathcal{N}_{i}$.
The set of \emph{neighbors} of $i$ consists of the particles $j$
whose neighborhood regions intersect $\mathcal{N}_{i}$, $\mathcal{N}_{i}\cap\mathcal{N}_{j}\neq0$,
and which are of a species interacting with the species of particle
$i$, i.e., $\mathcal{I}_{s_{i}s_{j}}=\mathcal{T}$.

We will assume that when a particle does not interact with other particles
we can easily follow its time evolution (motion), that is, given the
current configuration $\V{q}_{i}(t)$, we can probabilistically determine
the position at a later time $\V{q}_{i}(t+\D{t})$. This is a \emph{single-particle
problem} and can typically be solved analytically. For example, in
MD the particle trajectory is a unique (i.e., deterministic) straight
path, $\V{r}_{i}(t+\D{t})=\V{r}_{i}(t)+\V{v}_{i}\D{t}$, while in
diffusion problems it is the solution to a (stochastic) Langevin or
diffusion equation. Event-driven algorithms are efficient because
they use such analytic solutions to quickly propagate particles over
potentially large time steps as long as they are far enough from other
(interacting) particles. We will also assume that one can solve two-body
problems for the case when two particles are isolated from other particles
but may interact with each other. These two-body problems are typically
much more difficult to solve (quasi) analytically. Specific examples
will be given later.

\subsection{The Event Loop}

An AED algorithm consists of processing a sequence of time-ordered
events. Each particle $i$ must store some basic information needed
to predict and process events associated with it. The particle time
$t_{i}$ specifies the last time the configuration of particle $i$
was updated, $t_{i}\leq t$, where $t$ is the current simulation
time. Some particles may be time-driven and thus not have their own
event prediction. The rest of the particles are event-driven and each
such particle stores a prediction for its \emph{impending event} $\left(t_{e},p,\nu\right)$,
specified via the predicted time of occurrence $t_{e}$ (a floating
point number), the event \emph{partner} $p$ (an integer), and the
event \emph{qualifier} (type of event) $\nu$ (also an integer). The
partner $p$ could be some pre-specified invalid value to identify
time-driven particles. Note that the \emph{event schedules} must be
kept symmetric at all times, that is, if particle $i$ has an impending
event with $j$, then particle $j$ must have an impending event with
$i$. A particle may store multiple event predictions, in order to
avoid re-predicting events if the impending event is invalidated,
however, we will not explicitly handle this possibility due to the
complications it introduces.

The exact interpretation of $p$ and $\nu$, for a given particle
$i$, is application- and geometry-specific. Some common types of
events can be pre-specified by reserving certain values of the event
partner $p$, for example, we have used the following list for the
set of models presented here:

\begin{description}
\item [{$p=0$}] An \emph{update} of the event prediction for $i$, not
requiring an update of $\V{q}_{i}$. The value $\nu=0$ means that
$\V{q}_{i}$ has not changed since the last prediction for $i$ (thus
allowing stored information from previous predictions to be reused
if needed), $\nu=1$ means that an event occurred which did not alter
the geometry (for example, the position of $i$ is the same but its
velocity changed), while $\nu=-1$ means that this particle was just
inserted into the system and a geometry update is necessary as well.
\item [{$p=i>0$}] A single-particle event that requires an update of $\V{q}_{i}$.
The special value $\nu=0$ denotes a simple time advance of $i$ without
any additional event processing, $\nu<0$ denotes an event that does
not change the geometry (for example, only the velocity of a particle
changes), and $\nu>0$ is used for additional AS events that may also
change the geometry (e.g., particle decay).
\item [{$1\leq p\leq N$~and~$p\neq i$}] An unprocessed \emph{binary
reaction} between particles $i$ and $j=p$, with additional AS information
about the type of reaction stored in $\nu$, for example, elastic
collision, a certain chemical reaction, etc.
\item [{$p=\infty$}] A {}``boundary'' event requiring the update of
the particle geometry. If $\nu=0$ then only the protective region
$\mathcal{P}_{i}$ needs to be updated, if $\nu=-1$ then the neighborhood
$\mathcal{N}_{i}$ needs to be updated (collision with a \emph{virtual
boundary}), $\nu<-1$ denotes collisions with pre-specified boundaries
(such as hard walls), and $\nu>0$ specify AS boundary events (such
as collisions with reactive surfaces).
\item [{$p=-\infty$}] Denotes an invalid event, meaning that this particle
is not in the event queue and is handled separately, for example,
it is time-driven.
\end{description}
It is important to point out that we are not suggesting that an actual
implementation needs to use integers to identify different types of
events. In an object-oriented framework events may be represented
as objects that inherit from a base event class and have methods to
handle them, with the base implementation providing handlers for certain
pre-defined (single, pair, and boundary) types of events. We do not
discuss here the possible ways to organize an inheritance hierarchy
of classes for AED simulations, since such a hierarchy involves multiple
complex components, notably a module for handling boundary conditions
in static and dynamic environments, a module for handling static and
dynamic particle geometry (overlap, neighborhoods, neighbor searches,
etc.), an event-dispatcher, a visualization module, application-specific
modules for event scheduling and handling, etc.

Algorithm \ref{ProcessEvent} represents the main \emph{event loop}
in the AED algorithm, which processes events one after the other in
the order they occur and advances the global time $t$ accordingly.
It uses a collection of other auxiliary geometry-specific (GS) or
application-specific (AS) steps, as marked in the algorithm outline.
Specific examples of various GS and AS steps are given in the next
section.

\begin{Algorithm}{\label{ProcessEvent}Process the next event in
the event queue.}

\begin{enumerate}
\item \label{PopTopOfHeap}Find (query) the top of the event queue (usually
a heap) to find the next particle $i$ to have an event with $p$
at $t_{e}$. Note that steps marked as (AS+C) below may reorder the
queue and/or cycle back to this step.
\item \label{NextExternalEvent}Find the next {}``external'' event to
happen at time $t_{ex}$, possibly using an additional event queue
(AS).
\item \label{ProcessExternalEvent}If $t_{ex}<t_{e}$ then process the external
event (AS) and cycle back to step \ref{PopTopOfHeap}.
\item Remove $i$ from the event queue and advance the global simulation
time $t\leftarrow t_{e}$.
\item \label{CheckForOverlaps}If $p=0$ and $\nu=-1$ then build a new
$\mathcal{N}_{i}$ and then check if $i$ overlaps with any of its
new neighbors (GS). If it does, process the associated reactions (AS+C),
otherwise build a new $\mathcal{P}_{i}$ as in step \ref{UpdateProtectiveRegion}.
\item \label{SingleParticlePropagator}Else if $p=i$ then update the configuration
of particle $i$ to time $t$ using a single-particle propagator (AS),
and set $t_{i}\leftarrow t$. If $\nu\neq0$ then process the single-particle
event (AS+C). If $\nu>0$ then search for overlaps as in step \ref{CheckForOverlaps}.
\item \label{TwoParticlePropagator}Else if $1\leq p\leq N$ then update
the configuration of particles $i$ and $j=p$ using a two-particle
propagator (AS), set $t_{i}\leftarrow t$ and $t_{j}\leftarrow t$,
and then process the binary reaction between $i$ and $j$ (AS+C).
This may involve inserting particle $j$ back into the queue with
$t_{e}=t$, $p=0$, $\nu=0$.
\item \label{ProcessBoundaryEvents}Else if $p=\infty$, then update $\V{q}_{i}$
and $t_{i}$ as in step \ref{SingleParticlePropagator}. If $\nu>0$
then process the boundary event (AS+C), otherwise

\begin{enumerate}
\item \label{UpdateProtectiveRegion}If $\nu=0$ then update $\mathcal{P}_{i}$
(AS+GS), typically involving an iteration over the neighbors of $i$.
\item \label{UpdateBoundingCell}Else if $\nu=-1$ then update $\mathcal{N}_{i}$
and identify the new neighbors of particle $i$ (GS). 
\item \label{ProcessGeometryEvent}Else if $\nu<-1$ process the geometry-induced
boundary event (GS+AS).
\end{enumerate}
\item \label{PredictNextEvent}Predict a new $t_{e}$, $p$, and $\nu$
for particle $i$ by finding the minimal time among the possible events
listed below. Each successive search needs to only extend up to the
current minimum event time, and may return an incomplete prediction
$t_{e}>t$, $p=0$, $\nu=0$, where $t_{e}$ provides a lower bound
on the actual event time.

\begin{enumerate}
\item \label{PredictNextProtection}When particle $i$ leaves $\mathcal{P}_{i}$
or $\mathcal{N}_{i}$ (AS).
\item \label{PredictNextDecay}When particle $i$ undergoes a single-particle
event (AS).
\item \label{PredictNextCollision}When particle $i$ first reacts with
a neighbor $j$ (AS), as found by searching over all neighbors $j$
whose protective region $\mathcal{P}_{j}$ intersects $\mathcal{P}_{i}$
(GS). If a particle $j$ gives the current minimum event time, remove
it from the event queue. If such a particle $j$ has an event partner
that is another particle (third party) $k\neq i$, update the positions
of $j$ and $k$ using the two-particle propagator as in step \ref{TwoParticlePropagator},
invalidate $k$'s event prediction by setting its $t_{e}\leftarrow t$,
$p\leftarrow0$, $\nu\leftarrow0$, and update its position in the
event queue (alternatively, one may use lazy invalidation strategies).
\end{enumerate}
\item Insert particle $i$ back into the event heap with key $t_{e}$ and
go back to step \ref{PopTopOfHeap}.
\end{enumerate}
\end{Algorithm}

\subsubsection{Non-Particle Events}

In a variety of applications the majority of events can be associated
with a specific particle, and one can schedule one event per particle
in the event queue. However, sometimes there may be events that are
associated with a (possible large) group of particles, or events that
are not specifically associated with a particle. We consider these
non-particle events as application-specific {}``external'' events
in Algorithm \ref{ProcessEvent}.

An important example of such an event are \emph{time step events}.
Namely, some group of particles may not be propagated asynchronously
using the event queue, instead, the particles in the group may be
updated synchronously, for example, in regular time intervals $\D{t}$.
There may in fact be multiple such groups each with their own timestep,
for example, each species might have its own time step. Alternatively,
all or some of the particles may be updated in a time-stepped manner
and additional asynchronous events may be processed in-between the
time step events. An example is molecular dynamics in which time-driven
handling is combined with event-driven handling for the hard-core
collisions \citet{Mixed_Hard_Soft_MD,EventDriven_ParticleField}.
The time step events should also be ordered in time and the next one
chosen as the external event in Step \ref{NextExternalEvent} in Algorithm
\ref{ProcessEvent}. A separate priority queue may be used for ordering
the external events. In general, there may be an event queue of events
associated with particles, with cells, with species, etc. These may
be separate queues that are joined at the top or they can be merged
into a single heap.

\subsection{Near-Neighbor Search}

All large-scale particle-based algorithms use various geometric techniques
to make the number of neighbors of a given particle $O(1)$ instead
of $O(N)$. Reference \citet{Event_Driven_HE} provides extensive
details (and illustrations) of these techniques for hard spheres and
ellipsoids; here we summarize only the essential components.

\subsubsection{Linked List Cell (LLC) Method}

The most basic technique is the so-called \emph{linked list cell}
(LLC) method. The simulation domain, typically an orthogonal box,
is partitioned into $N_{c}$ cells, typically cubes. Each particle
$i$ stores the cell $c_{i}$ to which its centroid belongs, and each
cell $c$ stores a list $\mathcal{L}_{c}$ of all the particles it
contains (usually we also store the total number of particles in the
cell). Given a particle and a range of interaction, the lists of potential
neighbors is determined by scanning through the neighboring cells.
For maximal efficiency the cell should be larger than the largest
range of interaction so that only the nearest-neighbor cells need
to be searched. There are more sophisticated neighbor search methods
developed in the computational geometry community, such as using (colored)
quad/oct-trees, however, we are not aware of their use in AED implementations,
likely because of the implementation complexity. This is an important
subject for future research.

It is important to point out that in certain applications the cells
themselves play a crucial role in the algorithm, typically as a means
to provide mesoscopic averages of physical variables (averaged over
the particles in a given cell) used to switch from a particle-based
model to a continuum description. For example, in PIC (particle-in-cell)
algorithms for plasma simulation, the cells are used to solve for
background electric fields using FFT transforms \citet{AED_PIC}.
In DSMC, the algorithm stochastically collides pairs of particles
that are in the same cell. In some applications, events may be associated
with the cells themselves, instead of or in addition to the events
associated with particles \citet{IntraCellSignaling_BKL}. Usually
the same cells are used for both neighbor searches and averaging for
simplicity, however, this may not be the optimal choice in terms of
efficiency.

For a method that only uses the LLC method for neighbor searches,
the neighborhood region $\mathcal{N}_{i}$ is composed of the (typically
$3^{d}$, where $d$ is the dimensionality) cells that neighbor $c_{i}$,
including $c_{i}$ itself. The protection region $\mathcal{P}_{i}$
may be a simple geometric region like a sphere inscribed in $\mathcal{N}_{i}$
(sphere of diameter smaller than the cell size), it may be that $\mathcal{P}_{i}\equiv c_{i}+\mathcal{C}_{i}$,
or maybe $\mathcal{P}_{i}\equiv\mathcal{N}_{i}$.

\subsubsection{Near-Neighbor List (NNL) Method}

Another neighbor search method is the \emph{near-neighbor list} (NNL)
method, which is described for hard particles in Ref. \citet{Event_Driven_HE}.
The idea is to use as $\mathcal{N}_{i}$ a region that (when it is
created) is just an enlargement of the particle by a certain scaling
factor $\mu>1$. When $\mathcal{N}_{i}$ is created the method also
creates a (linked) list of all the neighborhoods that intersect it,
to form $\mbox{NNL}(i)$ (hard walls or other boundaries may also
be near neighbors). This list of (potential) \emph{interactions} can
then be reused until the particle core $\mathcal{C}_{i}$ protrudes
outside of $\mathcal{N}_{i}$. This reuse leads to great savings in
situations where particles are fairly localized.

Note that the LLC method is still used in order to create $\mathcal{N}_{i}$
and $\mbox{NNL}(i)$ even if NNLs are used, in order to keep the maximal
cost of pairwise searches at $O(N)$ instead of $O(N^{2})$. In some
situations (such as mixed MD/DSMC simulations as we describe later)
one may use NNLs only for a subset of the particles and use the more
traditional LLCs for others. In this case one can use $\mathcal{P}_{i}\equiv\mathcal{N}_{i}\cap(c_{i}+\mathcal{C}_{i})$
in order to ensure that both the NNLs (for those particles that have
them) and the LLCs (for all particles) are valid neighbor search methods.

\subsubsection{Cell Bitmasks}

Efficient handling of spatial information is an essential component
of realistic AED algorithms. For example, further improvements to
the basic LLC method may be required for certain applications in order
to maximize the efficiency of neighbor searches. The handling of boundary
conditions or domain-decomposition is an application-specific component
that is in some sense disjoint from the basic AED framework presented
here, however, it is often very important in practice. In this section,
which may be skipped at first reading, we describe an enhancement
to the basic LLC method that we have found very useful in handling
spatial information. 

In our implementation, in addition to the list of particles $\mathcal{L}_{c}$,
each cell $c$ stores a \emph{bitmask} $\mathcal{M}_{c}$ consisting
of $N_{\mbox{bits}}>N_{s}$ bits (bitfields), where $N_{s}$ is the
number of species. These bits may be one (set) or zero (not set) to
indicate certain properties of the cell, specifically, what species
of particles the cell contains, whether the cell is event or time
driven, and to specify boundary conditions. Bit $\gamma$ in the bitmask
$\mathcal{M}_{c}$ is set if the cell $c$ contained a particle of
species $\gamma$ in the near past and may still contain a particle
of that species. The bit is set whenever a particle of species $\gamma$
is added to the cell, and it should be cleared periodically if the
cell no longer contains particles of that species. When performing
a neighbor search for a particle $i$, cells not containing particles
of species that interact with species $s_{i}$ are easily found (by
OR'ing the cell masks with the $s_{i}$'th row of $\mathcal{I}_{s_{i}s_{j}}$)
and are simply skipped. This can significantly speed up the neighbor
searches in cases where not all particles interact with all other
particles.

For the purposes of a combined event-driven with time-driven algorithm
one may also need to distinguish those cells where particles are treated
using a fully event-driven (ED) scheme. We use one of the bits in
the bitmasks, bit $\gamma_{ED}$, to mark \emph{event-driven (ED)
cells}, and the choice of such cells is in general application specific.
For example, for diffusion problems, cells with a high density may
be treated more efficiently using time-driven (small hopping) techniques
while areas of low density may be treated more efficiently using the
asynchronous event-driven algorithm. In our combined MD with DSMC
algorithm cells that contain or that are near non-DSMC particles are
event-driven while those containing only DSMC particles are treated
as time-driven cells. Note that all of the cell bitmasks should be
reset and then re-built (i.e., refreshed) periodically.

The cell masks can also be used to specify partitionings of the simulation
domain. This is very useful in specifying more general boundary conditions
in situations when the event-driven simulation is embedded inside
a larger domain that is not simulated explicitly. For example, a molecular
dynamics simulation may be embedded in a multiscale solver where the
surrounding space is treated using a continuum method (finite element
or finite volume, for example) coupled to the particle region through
an intermediary boundary layer. Similar considerations apply in parallelization
via domain decomposition, where the simulation domain simulated by
a single processing element (PE) or logical process (LP) is embedded
inside a larger region where other domains are simulated by other
PEs/LPs.

We classify the cells as being \emph{interior, boundary, and external
cells} (a specific illustrative example is given in Fig. \ref{TetheredPolymer.partitioning}).
Our implementation uses bits in the cell bitmasks to mark a cell as
being boundary (bit $\gamma_{B}$), or external (bit $\gamma_{E}$),
the rest are interior. Interior cells are those for which complete
boundary conditions are specified and that cannot be directly affected
by events occurring outside of the simulation domain (the interior
cells are divided into event-driven and time-driven as discussed earlier).
Boundary cells surround the interior cells with a layer of cells of
thickness $w_{B}^{I}\geq1$ cells (typically $w_{B}^{I}=1$ or $w_{B}^{I}=2$)
and they represent cells that are affected by \emph{external events}
(i.e., events not simulated directly). External cells are non-interior
cells that are not explicitly simulated, rather, they provide a boundary
condition/padding around the interior and boundary cells. This layer
must be at least $w_{B}^{E}$ cells thick, and the cells within a
layer of $w_{B}^{E}$ cells around the simulation domain (interior
together with boundary cells) are marked as both external and boundary
cells (e.g., these could be ghost cells in parallel simulation). It
is often the case that $w_{B}^{I}=w_{B}^{E}=w_{B}$. All of the remaining
cells are purely external cells and simply ignored by the simulation.

\section{Model Examples}

In this section we present the handling of the various AS and GS steps
in Algorithm \ref{ProcessEvent} for three specific model applications.

\subsection{Event-Driven Molecular Dynamics (EDMD)}

Hard-particle molecular dynamics is one of the first applications
of AED algorithms in computational science, and is discussed in more
detail in Ref. \citet{Event_Driven_HE} and references therein. Hard-sphere
MD has been used extensively in simulations of physical systems over
the past decades and is also one of the few particle AED algorithm
that has been studied in the discrete-event simulation community,
as the \emph{billiards} problem. Because of this rich history and
extensive literature we only briefly discuss EDMD focusing on how
it fits our general framework.

The basic type of event in EDMD are \emph{binary collisions}, which
alter the momenta of two touching (and approaching) particles, typically
based elastic collision laws (conservation of momentum and energy).
Collisions are assumed to have no duration and (very unlikely) triple
collisions are broken up into a sequence of binary ones. In-between
collisions particles move ballistically along simple trajectories
such as straight lines (force-free motion) or parabolas (constant-acceleration
motion). For hard spheres, event time predictions are based on (algebraic)
methods for finding the first root of a polynomial equation (linear,
quadratic or quartic \citet{Mixed_Hard_Soft_MD}). For particle shapes
that include orientational degrees of freedom, such as ellipsoids,
numerical root finding techniques need to be used \citet{Event_Driven_HE}.

\subsubsection{AED Implementation of EDMD}

When LLCs are used, the main type of boundary event are \emph{cell
transfers}, which occur when the centroid of a particle $i$ collides
with the boundary of the cell $c_{i}$. If the cell is at the boundary,
a \emph{unit cell change} occurs for periodic boundaries (i.e., wrapping
around the torus), and \emph{hard-wall collisions} occur for hard-wall
boundaries. When NNLs are used, cell transfers do not have to be processed
(i.e., one can set $\mathcal{P}_{i}\equiv\mathcal{N}_{i}$), unless
it is important to have accurate LLCs {[}as in DSMC, where $\mathcal{P}_{i}\equiv\mathcal{N}_{i}\cap(c_{i}+\mathcal{C}_{i})$].

The main AS steps in Algorithm \ref{ProcessEvent} for (classical)
MD simulations are:

\begin{itemize}
\item Step \ref{SingleParticlePropagator} consists of updating the particle
position, and possibly also velocity, ballistically.
\item Step \ref{TwoParticlePropagator} consists of updating the positions
of each of the particles separately, as in step \ref{SingleParticlePropagator},
and then updating their velocities taking into account the collisional
exchange of momentum.
\item For collisions with hard walls in Step \ref{ProcessBoundaryEvents},
the particle velocity is updated accordingly. Cell transfers in step
\ref{ProcessGeometryEvent} consist of updating the LLCs by removing
the particle from its current cell and inserting it into its new cell
(found based on the direction of motion of the centroid). If the new
cell is across a periodic boundary, the centroid is translated by
the appropriate lattice cell vector (if NNLs are used, this may require
updating information relating to periodic boundaries for each interaction).
\item Step \ref{PredictNextEvent} is the most involved and time consuming:

\begin{itemize}
\item In step \ref{PredictNextProtection}, the time of the next cell transfer
is predicted based on the centroid velocity. If NNLs are used, then
the time of (virtual) collision of the core $\mathcal{C}_{i}$ with
the interior wall of $\mathcal{N}_{i}$ is also calculated.
\item In step \ref{PredictNextCollision}, predictions are made for binary
collisions between particle $i$ and each of the particles in neighboring
cells, or between $i$ and each of the particles in $\mbox{NNL}(i)$.
\end{itemize}
\end{itemize}

\subsection{Stochastic Event-Driven Molecular Dynamics (SEDMD)}

We have recently developed a novel AED algorithm for simulating hydrodynamics
at the molecular level that combines DSMC, which is a method for simulating
hydrodynamic transport, with event-driven molecular dynamics (EDMD).
The algorithm replaces some of the deterministic collisions in EDMD
with stochastic collisions and we term it Stochastic EDMD (SEDMD).
The algorithm is described in more detail in Ref. \citet{DSMC_AED}.

First we describe how to transform the traditional time-driven DSMC
algorithm \citet{DSMCReview_Garcia} into an event-driven algorithm,
and then combine this algorithm with EDMD. Finally, we explain how
to achieve higher efficiency by reverting the DSMC component to time-driven
handling.

\subsubsection{DSMC for Hydrodynamics}

The DSMC algorithm can be viewed as an approximation to molecular
dynamics in cases when the internal structure of the fluid, including
the true equation of state, is not important. In particular, this
is the case when simulating a solvent in applications such as the
simulation of large polymer chains in solution. The exact trajectories
of the solvent particles do not really matter, and what really matters
are the (long time and long range) hydrodynamic interactions that
arise because of local energy and momentum exchange (viscosity) and
conservation (Navier-Stokes equations). Any method that simulates
the correct momentum transfer localized at a sufficiently small scale
is a good replacement for full-scale MD, and can lead to great computational
savings when a large number of solvent molecules needs to be simulated.

DSMC \citet{DSMCReview_Garcia} achieves local momentum exchange and
conservation by performing a certain number of stochastic collisions
between randomly chosen pairs of particles that are inside the same
cell. The collision rate inside a cell containing $N_{L}$ particles
is proportional to $N_{L}(N_{L}-1)$ with a pre-factor that can be
based on theory or fitted to mimic that of the full MD simulation.
For hard spheres, the probability of choosing a particular pair $ij$
is proportional to the relative velocity $v_{ij}^{\mbox{rel}}$, and
typically a rejection technique (null-method technique) is used when
choosing pairs. Specifically, the collision rate is made proportional
to the maximal possible relative velocity $v_{\mbox{rel}}^{\mbox{\mbox{max}}}$,
and a randomly chosen pair $ij$ is rejected or accepted with probability
$v_{ij}^{\mbox{rel}}/v_{\mbox{rel}}^{\mbox{\mbox{max}}}$. A \emph{pair
rejection} involves a small calculation and a random number generation
and is thus rather inexpensive, as long as the acceptance probability
is not too small, which can typically easily be achieved by a judicious
(but still rigorous) choice for $v_{\mbox{rel}}^{\mbox{\mbox{max}}}$.
Alternative rules for selection of collision partners and for the
choice of the post-collisional velocities can give thermodynamically-consistent
non-ideal DSMC fluids \citet{SHSD_PRL}.

Traditionally DSMC is performed using a time-driven approach: The
particles are first propagated ballistically by a certain time step
$\D{t}$ and then sorted into cells accordingly, and then an appropriate
number of stochastic collisions are carried out in each cell. Furthermore,
the existence of a finite step leads to errors in the transport coefficients
(such as shear viscosity) of order $\D{t}^{2}$ \citet{DSMC_TimeStepError},
in addition to the errors inherent in DSMC that are of order $\D{x}^{2}$
\citet{DSMC_CellSizeError}, where $\D{x}$ is the size of the cells.
Therefore, in traditional DSMC $\D{t}$ should be small enough so
that particles move only a fraction of the cell size and a fraction
of the mean free path during one step \citet{DSMC_CellSizeError,DSMC_TimeStepError}.
As a consequence, only a fraction (10-25\%) of the particles actually
undergo a collision, and the rest of the particles are propagated
needlessly. We note, however, that DSMC can be extended into the dense-fluid
regime in which case the collisional frequency is high and thus most
particles undergo a stochastic collision at every time-step \citet{SHSD_PRL}.

\subsubsection{AED Implementation of DSMC}

An alternative event-driven implementation of DSMC explicitly predicts
and process cell transfers, just as in EDMD algorithm. Particles positions
are thus only updated when needed, and there is no time step error
since it is easy to obtain and numerically evaluate closed-form expressions
for the time of the cell crossings. The DSMC particles are represented
as a species $\delta$ for which particles do not interact with other
DSMC particles (i.e., $\mathcal{I}_{\delta\delta}=\mathcal{F}$),
so that the MD algorithm does not predict binary collisions for the
DSMC particles. Instead, stochastic binary collisions are added as
an external Poisson event of the appropriate rate. One approach is
to maintain a global time-of-next-DSMC-collision $t_{sc}$ to determine
when a stochastic collision is attempted, and to use \emph{cell rejection}
to select a host cell for the collision. The rate of DSMC collisions
is chosen according to the cell with maximal occupancy $N_{L}^{\mbox{max}}$,
and a randomly chosen cell of occupancy $N_{L}$ is accepted with
probability $N_{L}(N_{L}-1)/\left[N_{L}^{\mbox{max}}(N_{L}^{\mbox{max}}-1)\right]$.
It is possible to avoid cell rejections altogether, using any of the
multiple {}``rejection-free'' techniques common to KMC simulations
\citet{AdvancedKMC_Tutorial}. For example, the cells could be grouped
in lists based on their occupancy and then an occupancy chosen first
(with the appropriate weight), followed by selection of a cell with
that particular occupancy.

Most of the AS steps in Algorithm \ref{ProcessEvent} for DSMC are
shared with MD. The different steps are associated with the processing
of stochastic collisions:

\begin{itemize}
\item In Step \ref{NextExternalEvent} $t_{ex}$ is the time of the the
next DSMC collision. If a rejection-free technique is used the cell
at the top of the cell event queue is chosen, otherwise, a cell is
selected randomly and accepted or rejected based on its occupancy.
If the cell is rejected, no collisions are processed.
\item In Step \ref{ProcessExternalEvent} a pair of particles $i$ and $j$
is randomly selected from the previously-chosen cell, and accepted
or rejected for collision based on the relative velocity. If accepted,
a randomly-chosen amount of momentum and energy is exchanged among
the particles and they are moved to the top of the event queue with
$t_{e}=t$, $p=0$, and $\nu=1$.
\end{itemize}
We have validated our event-driven DSMC algorithm by comparing against
published results for plane Poiseuille flow of a rare gas obtained
using traditional time-driven DSMC \citet{DSMC_PlanePoiseuille}.

\subsubsection{Stochastic Event-Driven Molecular Dynamics (SEDMD)}

The event-driven DSMC algorithm has few advantages over a time-driven
approach, which are outweighed by the (implementation and run-time)
cost of the increased algorithmic complexity. However, the AED variant
of DSMC is very similar to EDMD, and therefore it is relatively simple
to combine DSMC with EDMD in an event-driven framework. This enables
the simulation of systems such as colloids or hard-sphere bead-chain
polymers \citet{EDMD_Polymers_Hall} in solution, where the solute
particles are treated using MD, and the less-important solvent particles
are treated approximately using DSMC. The solvent-solute interaction
is still treated with MD. Similar studies have already been carried
out using time-driven MD for the solute particles and a simplified
variant of DSMC that approximates the solute particles as point particles
and employs multi-particle stochastic collisions \citet{DSMC_MPCD_MD_Kapral}.

We have designed and implemented such a combined algorithm, which
we term Stochastic Event-Driven Molecular Dynamics (SEDMD) \citet{DSMC_AED}.
The implementation is almost identical to classical hard-sphere MD,
with the addition of a new \emph{DSMC species} $\delta$ for which
particles do not interact with particles of the same species (i.e.,
$\mathcal{I}_{\delta\delta}=\mathcal{F}$). That is, the DSMC {}``hard
spheres'' freely interpenetrate each other, but collide as usual
with other species. An external \emph{stochastic collision} event
occurring as a Poisson process of the appropriate rate is used to
collide randomly chosen pairs of nearby DSMC particles, as described
in the previous section.

The algorithm is much more efficient than EDMD because of the replacement
of deterministic collisions with stochastic ones; however, it is still
not as efficient as classical time-driven DSMC, especially at higher
collision rates. The two main causes of this are the overhead of the
event queue operations in the AED variant of DSMC (note that the queue
needs to be updated after every stochastic collision), and also the
cost of neighbor searches. Namely, for each DSMC particle the nearby
cells need to be searched in Step \ref{PredictNextCollision} to make
sure they do not contain any solute particles (recall that cell bitmasks
are used to efficiently implement this). In cases when most of the
cells contain only DSMC particles, this can introduce significant
overheads.

This inefficiency can be corrected by combining time-driven with event-driven
handling. Specifically, only those those cells that neighbor cells
that contain non-DSMC particles are marked as event-driven (ED), the
rest are time-driven (TD). DSMC particles outside the marked region
are treated more efficiently, using time-stepping and without any
neighbor searches, while the DSMC and MD particles inside the marked
region are treated as in MD (with the addition of stochastic collisions
among DSMC particles). In cases when the solute particles are much
larger than the solvent particles, NNLs are used with a special technique
called bounded sphere-complexes \citet{Event_Driven_HE} to handle
neighbor searches for the large particles. Whenever DSMC particles
transfer from a TD to an ED cell they are inserted into the event
queue with an immediate update event, and if needed their neighborhood
region and NNL is constructed. Similarly, when DSMC particles transfers
from an ED to a TD cell they are removed from the event queue and
their neighborhood region and NNL is destroyed.

The time-driven particles are updated together with a fixed time step
$\D{t}$. These \emph{time step events} are treated as a special external
event and thus processed in correct time order with the events scheduled
for the particles in the combined MD/DSMC region. Stochastic collisions
are only processed at time step events, exactly as in traditional
DSMC algorithms.

\subsubsection{Open Boundary Conditions}

In simulations of polymer chains in solution in three dimensions,
a very large number of solvent particles is required to fill the simulation
domain. The majority of these particles are far from the polymer chain
and they are unlikely to significantly impact or be impacted by the
motion of the polymer chain. These particles do not need to be simulated
explicitly, rather, we can think of the polymer chain and the surrounding
DSMC fluid as being embedded in an infinite reservoir of DSMC particles
which enter and leave the simulation domain following the appropriate
distributions.

Such \emph{open (Grand Canonical) boundary conditions} (BCs) are often
used in multi-scale (coupled) simulations. The combination of a partially
time-driven algorithm and an unstructured (ideal gas) DSMC fluid makes
it very easy to implement open BCs by inserting DSMC particles in
the cells surrounding the simulation domain only at time-step events,
based on very simple distributions. At the beginning of a time step
event, after possibly rebuilding the cell masks, the time-driven DSMC
particles are propagated as usual. If there are external cells, (trial)
\emph{reservoir particles} are then inserted into the cells that are
both external and boundary. The trial particles are thought to be
at a time $t-\D{t}$, and are propagated by a time step $\D{t}$ to
the current simulation time. Only those particles that move into a
non-external cell are accepted and converted into real particles.
Following the insertion of reservoir particles stochastic collisions
are processed in each cell as usual.

\begin{figure*}
\includegraphics[width=0.4\paperwidth,keepaspectratio]{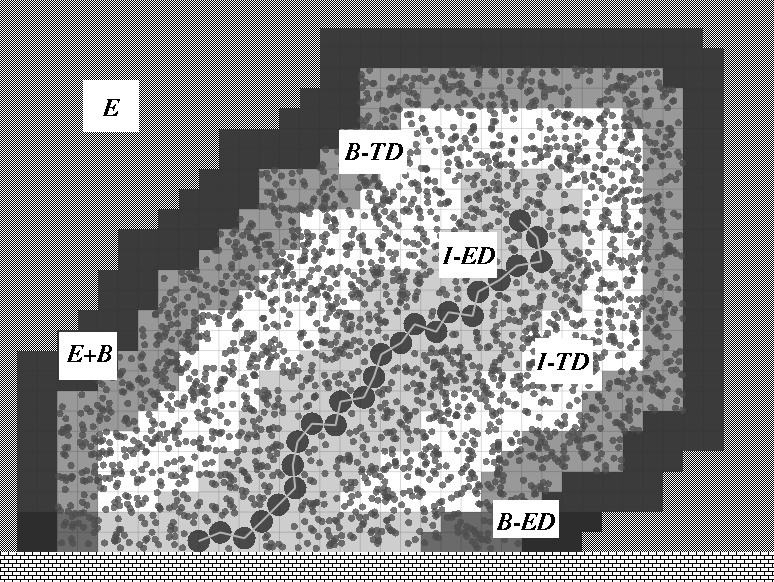}\hspace{0.5cm}\includegraphics[width=0.3\paperwidth,keepaspectratio]{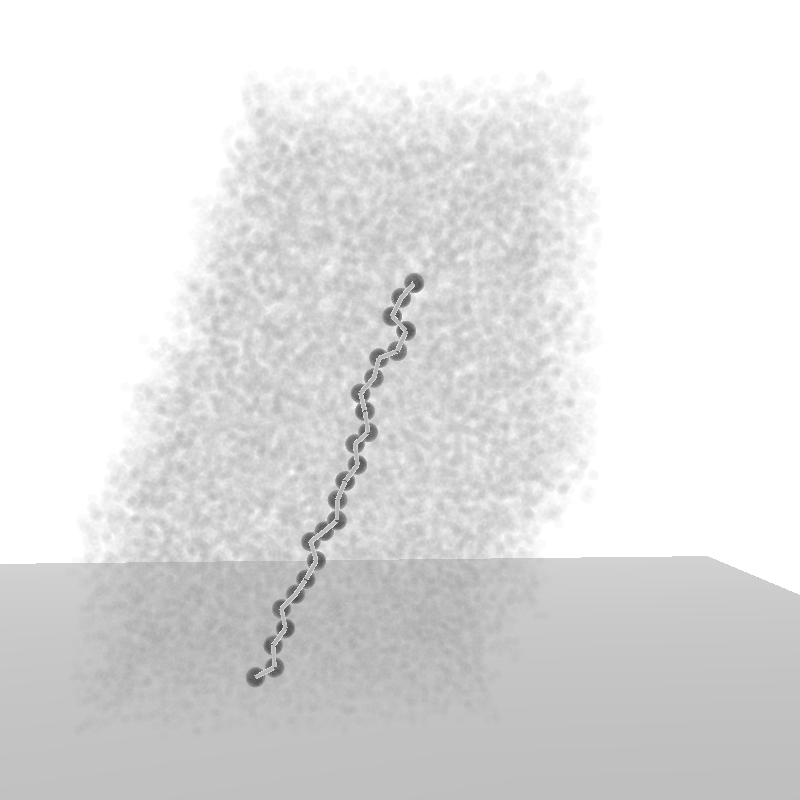}

\caption{\label{TetheredPolymer.partitioning}The partitioning of the domain
into interior (I) {[}either event-driven (ED) or time-driven (TD)],
boundary (B), and external (E) cells in two (left) and three (right)
dimensions for a polymer chain of $25$ beads tethered to a hard wall.
The cells are shaded in different shades of gray and labeled in the
two-dimensional illustration. The DSMC particles are also shown.}
\end{figure*}

Figure \ref{TetheredPolymer.partitioning} provides an illustration
of the division (masking) of the cells for the simulation of a tethered
polymer in two and three dimensions. The division of the cells into
event-driven, interior, boundary and external cells is rebuilt periodically
during the simulation. This rebuilding may only happen at the beginning
of time steps, and requires a synchronization of all of the particles
to the current simulation time, a complete rebuilding of the cell
bitmasks, and finally, a re-initialization of the event processing.
Importantly, particles that are in purely external cells are removed
from the simulation and those that are in event-driven cells are re-inserted
into the event queue scheduled for an immediate update event.

\subsection{First-Passage Kinetic Monte Carlo (FPKMC)}

An exact AED algorithm for kinetic Monte Carlo (KMC) simulation of
a collection of diffusing particles was recently proposed in Ref.
\citet{FPKMC_PRL}. The main difference with MD is that the equation
of motion of the particles is not a deterministic but rather a stochastic
ordinary differential equations (ODE). Additionally, reactions between
particles lead to new types of events such as particles changing species,
appearing or disappearing. This complicates the implementation but
is conceptually simple to incorporate into the basic AED algorithmic
framework.

Assuming that all of the required propagators can be obtained in closed
form and evaluated to within numerical precision, the FPKMC algorithm
is exact in the sense that it simulates a trajectory that is sampled
from the correct probability distribution. For pure diffusion with
transport coefficient $D$, the probability $c(\D{\V{r}},\D{t})$
of finding an isolated point particle at a displacement $\D{\V{r}}$
at time $\D{t}$ is the Green's function for the (time-dependent)
diffusion equation, $\partial_{t}c=D\grad^{2}c$, with no additional
boundary conditions. In three dimensions\[
c(\D{\V{r}},\D{t})=\left(4\pi\D{t}\right)^{-1/2}\exp\left[-\frac{\D{r}^{2}}{4D\D{t}}\right].\]
Particles that have a finite extent, such as spheres and cubes, are
easily handled by considering their centroids as the diffusing point
particles. Particles with orientational degrees of freedom are more
difficult to handle and for now we focus on the sphere case.

Assume that the protective region $\mathcal{P}_{i}$ is disjoint from
all other protective regions and the core $\mathcal{C}_{i}$ is restricted
to remain within $\mathcal{P}_{i}$. For point particles, one can
show that the probability $c(\D{\V{r}},\D{t})$, \emph{conditional}
on the fact that the particle never leaves the interior of $\mathcal{P}_{i}$,
is again a Green's function for the diffusion equation but with the
additional boundary condition that $c$ vanish on the boundary of
$\mathcal{P}_{i}$. A single-particle propagator consists of \emph{sampling}
from such a probability distribution $c_{1}$. For simple protective
regions such as cubes or spheres relatively simple closed-form solutions
for $c_{1}$ exist. The probability distribution $c_{1}$ is only
valid under the assumption that a given particle $i$ remains inside
$\mathcal{P}_{i}$. From $c_{1}$ (specifically, the flux $\grad c_{1}$
on the boundary of $\mathcal{P}_{i}$) one can also find the probability
distribution that a particle first leaves $\mathcal{P}_{i}$ for the
first time at a time $\tilde{t}$ and at position $\tilde{\V{r}}$,
i.e., the \emph{first-passage} probability $J_{1}(\tilde{t},\tilde{\V{r}})$.
This distribution can be used to sample a time at which particle $i$
is propagated to the surface of $\mathcal{P}_{i}$, and then $\mathcal{P}_{i}$
is updated.

The basic idea of the First-Passage Kinetic Monte Carlo (FPKMC) algorithm
is to protect the particles with disjoint protective regions (an unprotected
particle has $\mathcal{P}_{i}\equiv\mathcal{C}_{i}$) and then use
single-particle propagators to evolve the system. Typically the protective
regions would have the same position and shape as the particle itself
but be enlarged by a certain scaling factor $\mu_{\mathcal{P}}>1$.
At some point in time, however, two particles $i$ and $j$ will collide
and thus cannot be protected with disjoint regions. Such nearly-colliding
pairs are protected by a \emph{pair protection region} $\mathcal{P}_{ij}$
(e.g., two intersecting spheres, each centered around one of the particles).
A \emph{pair propagator} $c_{2}(\D{\V{r}}_{i},\D{\V{r}}_{j},\D{t})$
is used to either find the first-passage time, that is, the time when
one of the particles leaves $\mathcal{P}_{ij}$, or to propagate the
pair conditional on the fact that both particles remain inside $\mathcal{P}_{ij}$.
Analytical solutions can be found by splitting the problem into independent
diffusion problems for the center of (diffusional) mass $\V{r}_{ij}^{(CM)}=(\D{\V{r}}_{i}+\D{\V{r}}_{j})/2$
and for the difference $\V{r}_{ij}^{(D)}=\D{\V{r}}_{i}-\D{\V{r}}_{j}$
walker (with some additional weighting factors for unequal particles).
This makes the pair propagator $c_{2}=c_{2}^{(D)}c_{2}^{(CM)}$ a
combination of two independent propagators, $c_{2}^{(D)}$ and $c_{2}^{(CM)}$,which
are simpler to solve for analytically. The difference-propagator is
typically more difficult to obtain analytically because the condition
for collision $d_{ij}=D_{ij}$ forms an additional absorbing boundary
for the difference walker, i.e., a collision occurs whenever the first-passage
propagator $J_{2}^{(D)}(\tilde{\V{r}}_{ij}^{(D)},\tilde{t}^{(D)})$
samples a point on that boundary. For repulsive particles \citet{EventDriven_Colloids}
the boundary $d_{ij}=D_{ij}$ would be reflective (zero-flux) instead
of absorbing.

If the particles are cubes closed-form solutions can easily be found
for the pair propagators \citet{FPKMC_PRL}, however, in general,
two-body propagators are considerably more complex (and thus costly)
than single-body ones. One can in fact use approximate numerical propagators
in which the difference walker takes small hops until it gets absorbed
at one of the boundaries of its protection. This is similar to how
one numerically predicts pair collisions between non-spherical particles
in MD simulations using time-stepping for just a single pair of particles
\citet{Event_Driven_HE}. We have implemented hopping pair propagators
for spheres. The only difference with the analytical propagators is
that the hopping trajectory may need to be reversed later if a third
particle forces a destruction of the pair protection before the scheduled
pair event. We have used a state-saving mechanism in which the state
of the random number generator at the beginning of the pair event
is saved and later restored if the predicted event is canceled. If
the predicted event does in fact occur we avoid the cost of repeating
the same hopping trajectory by saving the final state of the walk
(i.e., the position of the difference walker) when predicting pair
events.

\subsubsection{AED Implementation of FPKMC}

Geometric near-neighbor searches are an essential component of the
FPKMC algorithm, and the same methods (LLCs and NNLs) as in MD are
used. Cell transfers are not explicitly predicted or processed in
this algorithm, rather, whenever the position of a particle is updated
the LLCs need to be updated accordingly. When NNLs are used, the collision
of $\mathcal{C}_{i}$ with $\mathcal{N}_{i}$ may be sampled exactly,
or, alternatively, the neighborhood $\mathcal{N}_{i}$ may be updated
whenever $\mathcal{C}_{i}$ is very close to the inner wall of $\mathcal{N}_{i}$.
We say that a particle $i$ is \emph{protected against} particle $j$
or pair $jk$ if $\mathcal{P}_{i}$ is disjoint from $\mathcal{P}_{j}$
or $\mathcal{P}_{jk}$, similarly for pairs of particles. The goal
of neighbor searches is to protect a particle $i$ against other particles
and pairs with the largest possible $\mathcal{P}_{i}$. There is a
balance between rebuilding protective and neighborhood regions too
often and propagating the particles over smaller steps, and some experimentation
is needed to optimize the algorithm and minimize the number of neighbor
searches that need to be performed. Whenever a protection $\mathcal{P}_{i}$
is destroyed, particle $i$ should be inserted back into the event
queue with $t_{e}=t$, $p=\infty$, $\nu=0$, so that it is protected
again right away.

The main AS steps in Algorithm \ref{ProcessEvent} for FPKMC simulations
are:

\begin{itemize}
\item Steps \ref{NextExternalEvent} and \ref{ProcessExternalEvent} may
involve the processing of particle birth processes, where a particle
of a given species is introduced into the system to model external
fluxes. These are typically assumed to occur as a Poisson process
and therefore the time to the next birth is simply an exponentially
distributed number, with the total birth rate given as the sum of
the birth rates for each of the species. The birth process may be
spatially homogeneous or the rate may depend on the cell in which
the birth occurs. The newborn particles are inserted into the queue
with $p=0$, $\nu=-1$.
\item Step \ref{SingleParticlePropagator} consists of sampling $\V{r}_{i}$
from $c_{1}$ or $J_{1}$ and typically also rebuilding $\mathcal{P}_{i}$
as in step \ref{UpdateProtectiveRegion}. For $\nu>0$, a particle
decay reaction may be processed, i.e., particle $i$ may disappear
to produce zero or more {}``product'' particles, which are inserted
into the queue with $p=0$, $\nu=-1$.
\item Step \ref{TwoParticlePropagator} consists of sampling positions $\V{r}_{i}$
and $\V{r}_{j}$ from the appropriate distribution:

\begin{itemize}
\item If the event is the decay of $i$ or $j$, then $c_{2}^{(D)}$ and
$c_{2}^{(CM)}$ are sampled, and then the decay reaction is processed.
\item If the event is the collision of $i$ and $j$, then $c_{2}^{(CM)}$
and $J_{2}^{(D)}$ are sampled, and the appropriate reaction (e.g.,
annihilation, coalescence, chemical reaction, etc.) is processed.
This may destroy $i$ and/or $j$ and/or create new particles to be
inserted into the queue (with $p=0$, $\nu=-1$).
\item If the event is the dissolution of the pair $ij$, then either $c_{2}^{(CM)}$
and $J_{2}^{(D)}$, or $c_{2}^{(D)}$ and $J_{2}^{(CM)}$ are sampled,
particle $j$ is inserted back into the queue with $p=0$, $\nu=0$,
and $\mathcal{P}_{i}$ is updated as in step \ref{UpdateProtectiveRegion}
(this may protect the particles $i$ and $j$ as a pair again).
\end{itemize}
\item Step \ref{UpdateProtectiveRegion} is the primary type of event in
step \ref{ProcessBoundaryEvents} and consists of updating $\mathcal{P}_{i}$.
The processing of such {}``virtual'' collisions with $\mathcal{P}_{i}$
consists of searching for the nearest protection region $\mathcal{P}_{j}$
or $\mathcal{P}_{jk}$ among the neighbors of particle $i$ (either
using LLCs or NNLs). Particle $i$ is then protected against that
nearest neighbor. If this makes $\mathcal{P}_{i}$ too small then
particle $j$ or pair $jk$ is propagated to time $t$ and $\mathcal{P}_{j}$
or $\mathcal{P}_{jk}$ destroyed. Finally, $\mathcal{P}_{i}$, $\mathcal{P}_{ij}$
or $\mathcal{P}_{ik}$ is constructed again, depending on the exact
local geometry.
\item Step \ref{PredictNextEvent} consists of sampling event times from
the appropriate distributions:

\begin{itemize}
\item In step \ref{PredictNextProtection} $J_{1}$ is sampled.
\item In step \ref{PredictNextDecay} an exponentially distributed time
is generated based on the decay rates for species $s_{i}$.
\item In step \ref{PredictNextCollision} $J_{2}$ is sampled, as well as
a decay time for each of the two particles, and the earliest of the
three times is selected.
\end{itemize}
\end{itemize}

\subsubsection{Time-Driven Handling}

Under certain conditions event-driven handling of diffusion may become
inefficient or cumbersome. For example, since all protections are
either of single particles or pairs of particles, very small protective
regions need to be used in very dense clusters of particles where
there may be many nearly colliding triplets of particles. The hops
taken by the asynchronous algorithm will then become just as small
as what a time-driven (approximate) algorithm would use, and it will
therefore be more efficient to use time-stepping for those particles
(and thus avoid the cost of event queue operations and also simplify
overlap search). Additionally, some particles or boundaries (e.g.,
grain boundaries) may have complex shapes and thus their diffusion
or interactions with other particles may be difficult to treat analytically
(even for spherical particles it has proven that exact pair propagators
are difficult to implement). As another example, tightly-bound collections
of particles (clusters) may act as a single particle that has complex
internal structure and dynamics (relaxation).

Just as for the SEDMD algorithm, one can add a time-driven component
to the asynchronous event-driven FPKMC algorithm. Particles that are
time-driven do not need to be protected against each other, instead,
they may be unprotected or they can be protected only against event-driven
particles. Time-driven particles whose protective regions overlap
form a cluster and should be treated using a synchronous algorithm
(this cluster may include all time-driven particles). In diffusion
problems it is often the case that different species have widely differing
diffusion coefficients and therefore very different time-steps will
be appropriate for different species. To solve this problem, one can
use the $n$-fold (BKL) synchronous event-driven algorithm inside
each cluster. In this algorithm, at each event (hop) only particles
of a single species hop by a small but non-negligible distance and
the rest remain in place, and more mobile particles are hopped more
frequently (with the correct relative frequency) than the less mobile
ones.

Note that a cluster may freely take hops up to the time of the next
event in the queue without stopping and modifying the event queue,
since it is known that those hops will not be pre-emptied by another
event. In some situations this may improve efficiency by reducing
the number of heap operations and also increasing the locality of
the code by focusing multiple events on the same (cached) small group
of particles.

\section{Discussion}

In this section we focus on some of the difficulties in deploying
AED algorithms in the simulation of realistic systems. The best-known
difficulty is the parallelization of AED algorithms \citet{TimeWarp_Book},
which we do not discuss here due to space limitations. Instead, we
will focus on the difficulties that make even serial simulations challenging.
It is important to point out that for problems involving hard particles,
that is, particles interacting with discontinuous potentials, event-driven
approaches are the only exact algorithm. Time-driven algorithms always
make an error due to the finite size of the time step $\D{t}$, and
typically $\D{t}$ must be much smaller than the actual time step
between events in order to guarantee that no events are missed.

The most involved aspect of implementing an AED algorithm for a particular
problem is the need for analytic solutions for various one- and two-body
propagators or event predictions. For EDMD, the difficulty is with
predicting the time of collision of two moving hard particles. For
hard spheres, this can be done analytically relatively easily (but
numerical care must be taken \citet{Mixed_Hard_Soft_MD}). When orientational
degrees of freedom are involved, however, a time stepped ODE-like
methodology is needed since analytical solutions are difficult to
obtain \citet{Event_Driven_HE}. This makes collision prediction much
more difficult to implement in a numerically-stable way and also much
more costly. In FPKMC, there is a need to analytically construct the
probability distributions $c_{1}$, $c_{2}$, $J_{1}$ and $J_{2}$,
or at least to find a way to efficiently sample from them. These distributions
are Green's functions for a time-dependent diffusion equation inside
regions such as spheres, cubes, spherical shells, intersection of
two spheres, or intersection of a cone and a sphere. For time-independent
problems such solutions can be constructed more easily, but for time-dependent
problems even the simple diffusion equation poses difficulties (analytical
solutions are typically infinite series of special functions in the
Laplace domain). Different boundary conditions such as reactive surfaces
require even more analytical solutions and tailor-made propagators.
The handling of more complex equations of motion such as the full
Langevin equation (which combines convection and diffusion) has not
even been attempted yet.

This makes designing a more general-purpose AED program virtually
impossible. This is to be contrasted to, for example, time-driven
MD where different interaction potentials can used with the same time
integrator. In general, time stepped approaches are the only known
way to solve problems for which analytical solutions do not exist,
including two-body problems in the case of EDMD or FPKMC. The algorithm
used in Ref. \citet{Event_Driven_HE} to predict the time of collision
for pairs of hard ellipsoids combines a time-driven approach with
the event-driven one. It does this without trying to combine them
in an intelligent way to avoid wasted computation (such as repeated
trial updates of the position of a given particle as each of its neighbors
is processed). We believe that such an intelligent combination will
not only provide a more general AED algorithm, but also make the algorithm
more robust numerically.

More generally, combining event-driven with time-driven algorithms
is important for efficiency (at a certain sacrifice in accuracy).
When the time step is large enough so that many events occur within
one time step one can use the event-driven algorithm in-between the
updates in the time-driven approach \citet{Mixed_Hard_Soft_MD,EventDriven_ParticleField,DefeasibleTimeStepping}.
When the time step is small, however, events occur sparingly only
in some of the time steps, and a different methodology is needed.
We proposed to add a new kind of \emph{time step event} that indicates
propagation over a small time step (e.g., for the set of particles
in pure DSMC cells). The essential advantage of event-driven algorithms
is that they automatically adjust to the time scale at hand, that
is, that they take the appropriate time step without any additional
input. The real challenge is to use time stepping in an event-driven
framework in which the time step is adjusted accordingly to not waste
computation, while still keeping the approximations controlled. We
have described such a framework for the SEDMD and FPKMC algorithms.
The merging of discrete-event and continuous-time models has also
been formalized in a rather general simulation framework \citet{MultiParadigmSimulation}.

Another unexplored or barely explored area is that of using controlled
approximations in AED algorithms. Approximate event-driven algorithms
have been used to handle a variety of processes, however, these often
use uncontrolled approximations. The approximate algorithms may reproduce
the required (macroscopic) physical averages just as well as the exact
algorithm would, however, controls are necessary to validate the simulations.
Examples of approximations that may be useful include ignoring unlikely
interactions between certain particles, approximate solutions instead
of exact propagators (such as expansions around the mean behavior),
etc. 

Almost all of the AED particle algorithms to date have focused on
single-particle or pair events. This is possible to do for hard particles
because exact triple collisions are extremely unlikely to occur. However,
for more realistic models, or when approximations are made, events
involving clusters of particles may need to be considered. For example,
a cluster of particles may evolve as a strongly-coupled (e.g., chemically
bonded) unit while interacting with other (e.g., freely diffusing)
single particles or clusters. An additional assumption in most AED
particle algorithms is that events affect only one or two particles,
so that the event predictions of the majority of particles remain
valid after processing an event. In some situations, however, there
may be global degrees of freedom and associated events that affect
all of the particles. For example, in MD there may be a macroscopic
strain rate that affects all of the particles, since all of the event
predictions are invalidated when the strain rate changes. In principle
the strain rate is coupled back to each of the particles, so that
every particle collision also changes the strain rate (albeit by a
small amount). In time-driven MD this is no problem since the evolution
of the system is synchronous and the strain rate evolves together
with the particles, however, in event-driven MD such coupling between
all of the particles makes it impossible to schedule events efficiently.
In this work we restricted our attention to problems with only short-range
interactions between the particles. However, many problems of interest
include long-range (e.g., electrostatic) interactions as well, and
these interactions effectively couple the motions of all of the particles. 

Finally, multi-algorithm and/or multi-scale combinations including
an AED component have not been explored to our knowledge. As an example,
consider the simulation of nano-structures during epitaxial film growth
\citet{KMC_SurfaceReactions}. At the smallest scales, time-driven
(first-principles or classical) MD is needed in order to study the
attachment, detachment, or hopping of individual particles or clusters.
Once the rates for these processes are known, lattice-based KMC can
be used to evolve the structure more quickly without simulating the
detailed (vibrational) motion of each atom. At larger scales, the
continuum-based FPKMC algorithm we described can be used to propagate
atoms over large distances in lower-density regions (across flat parts
of the surface). Finally, a time-driven continuum diffusion partial
differential equation solver can be used to model processes at macroscopic
length scales. Such ambitious investigations are a challenge for the
future.

\section{Acknowledgments}

This work performed under the auspices of the U.S. Department of Energy
by Lawrence Livermore National Laboratory under Contract DE-AC52-07NA27344
(UCRL-JRNL-234181).


\textbf{Aleksandar Donev} obtained a Ph.D. in Applied and Computational
Mathematics from Princeton University in 2006. He is presently a Lawrence
Postdoctoral Fellow at Lawrence Livermore National Laboratory.
\end{document}